# Spreadsheets in an ERP environment: not what the doctor ordered

No'am Newman
Doctoral candidate in Business Administration, Edinburgh Business School, Heriot Watt University
Kivunim.noam@gmail.com

*Abstract*
**Modern ERP systems contain flexible report generators but the tendency exists for users to export data to spreadsheets for manipulation, reporting and decision making. A purported reason for this is that some users are more familiar with personal reporting tools (spreadsheets) as opposed to enterprise reporting tools. The author's doctoral research intends to measure the extent of spreadsheet usage in ERP environments and to determine which factors facilitate this.**

## INTRODUCTION

The majority of spreadsheet research is concerned with the data contained *within* the spreadsheet; little attention has been paid to the *source* of those data. The SME business sector (on which my research is based) frequently bases its IT on spreadsheets, even though "for large spreadsheets, the issue is how many errors there are, not whether an error exists" [1].

In the past few decades, implementations of ERP systems in SMEs are becoming more and more frequent. Despite the flexible report generators which modern ERP systems contain, the tendency still exists for users to export data to spreadsheets for manipulation, reporting and decision making. Users also make use of *feral systems* (data contained in non-officially sanctioned programs) to store data which should be stored in the ERP system. Yen *et al.* note that "employees in the company often use spreadsheet applications as an alternative to replacing missing functions in ERP systems, but this leads to large amounts of manual work, which is time consuming, productivity impairing, and the creation of errors" [2].

SMEs are generally dynamic, frequently making new demands from their ERP systems. As modern ERP systems are extendable, allowing trained developers to add functionality (tables, screens, data manipulation routines, reports), such new demands can always be met from the ERP system. Advantages which ERP systems offer over spreadsheets are:
- Data locking on a per-record level
- Automatic backups
- Security (not everyone has permission to see everything)
- Audit trails
- Referential integrity
- Easy production of reports with different parameters
- Only one version – which is the most up-to-date version - of the data
- Procedures and reports are written (and debugged) by trained professionals

A personal anecdote from the company in which I work (which employs about 150 people), dating from 2012: the comptroller required about five work days to put together a series of monthly reports by means of spreadsheets. When these reports were implemented by the ERP system, the time required was reduced to a few hours. This simple example shows how using ERP can greatly reduce the time required to create complex reports when the data already resides within the ERP system.

There are cases where users of an ERP system do not trust the stored data and decide that they will maintain their own system and ensure its accuracy (feral systems). Whilst it is true that ERP depends on data which accurately reflects reality, and that inaccurate input will lead to inaccurate output (and thus to making wrong decisions), it would be better if these users spent their energy on working *within* the system (by improving the ERP data) so that all users can benefit from more accurate data, instead of keeping the accurate data to themselves.

## RESEARCH QUESTIONS AND HYPOTHESES

The research questions of my doctoral research are:
1. What is the prevalence of feral system usage in Israeli SMEs which have implemented ERP?
2. What are the factors which facilitate feral system usage in those companies?
3. What are the factors that influence an ERP user in using external reporting tools (feral systems) in preference to internal reporting tools (ERP)?
4. Do users who use feral systems have the necessary competency to do so?

The operational hypotheses are:
H1: there are specific factors which facilitate feral system usage in Israeli SMEs which have implemented ERP.
H2: there are specific factors which cause users to prefer external reporting tools to internal reporting tools.
H3: users who employ feral systems (spreadsheets) lack the required competency.

The research will concentrate on identifying factors leading to feral system usage. These factors can be divided into two classes: those dependent on the SME ('company factors') and personal factors.

The company factors include position within the ERP life cycle, business sector, number of licenses owned, degree of customisation, production approach and the degreee of the CEO's involvement in IT.

There are three common models of the ERP life cycle [3, 4, 5] whose main difference lies in their terminology. The research is primarily interested in companies which are in the 'evolution' phase of the ERP life cycle which is generally reached after about three years of using ERP. In this stage, advanced capabilities are integrated which provide benefits such as advanced planning and scheduling, supply-chain management and customer relationship management.

Four types of production approach have been identified [6]: make to stock, assemble to order, make to order and engineer to order. MTS and ATO items will be expected to have a well-defined bill of materials and so companies manufacturing with these approaches will be expected to use standard ERP techniques for production and costing. On the other hand, companies which manufacture with the MTO or ETO approaches may well decide that standard ERP techniques are not cost effective.

Personal factors can be divided into two categories: 'external" and 'internal'. The external factors are: age; years of experience with ERP, department, education, gender and mother tongue.

The internal factors include spreadsheet competency, spreadsheet self-efficacy, sense of ownership (with regard to the data), learning style and degree of satisfaction with ERP. The literature survey for these internal factors was primarily based on papers published in journals of applied psychology where the research was performed with users of computer systems. No papers were found which specifically researched psychological constructs with users of ERP systems, and so this doctoral research might be considered pioneering.

Another factor which might contribute to feral system usage is user training. Training is normally given in order to explain *how* to do something and not *why* to do something (another way of expressing this is that training generally deals with *technical competency* and not with *business context*) [7]. The average ERP user may be competent in the areas where training was given but may not be able to abstract the essence of the system in order to extrapolate new solutions.

Whilst ERP users will have received appropriate training, rare is the employee who has received formal training in spreadsheets; their knowledge is normally based on snippets of (not necessarily accurate or efficient) information which are passed from hand to hand. Employees who do use spreadsheets are generally unaware of professional standards of spreadsheet programming, which bodies such as EuSpRIG and SEMS try to disseminate.

Not all of the above factors are expected to influence feral system usage. The decision to adopt or ignore the ERP system is known as "post-adoptive behaviour" [8].

## METHODOLOGY

The research will be carried out by means of two separate questionnaires, one for companies and one for users. This division was created as it is unlikely that the average user will be aware of the company data. The user questionnaire has undergone several revision stages which resulted in the removal of extraneous questions and clarification of the remaining ones. This questionnaire now consists of 45 questions.

Possibly the most important conclusion reached from the preliminary pilot stage was that the questionnaire should be translated into Russian: many Russian immigrants are capable of working with ERP and talking in Hebrew, but their Hebrew reading skills are lacking which would reduce the response rate.

The pilot study in an SME using ERP is taking place at the time of writing, so whilst no data are currently available, preliminary results should be available by the time of presentation. In order to prevent bias, employees who work in the same company as the researcher will not participate in the final research.

With thanks to Professor P. O'Farrell and Dr F. Hermans.